\documentclass[]{pasj01}
\usepackage{lscape}
\usepackage{url}
\begin{document} 
\Received{}%{yyyy/mm/dd}
\Accepted{}%{yyyy/mm/dd}
%\Published{yyyy/mm/dd}

\title{MITSuME observation of V404 Cygni in the 2015 outburst: 
Two optical variable components with different variability}

%%% begin:list of authors
% Do NOT capitalize all letters in "textsc".
\author{Y. Tachibana\altaffilmark{1}, T. Yoshii\altaffilmark{1}, H. Hanayama\altaffilmark{2}, 
%Y. T. Tanaka\altaffilmark{3}, ....., 
and N. Kawai\altaffilmark{1}}
\email{tachibana@hp.phys.titech.ac.jp}
%% `\KeyWords{}' always has to be placed before `\maketitle'.

\altaffiltext{1}{Department of Physics, Tokyo Institute of Technology, 2-12-1, Ohokayama, Tokyo, Japan}
\altaffiltext{2}{Ishigakijima Astronomical Observatory, National Astronomical Observatory
of Japan, National Institutes of Natural Sciences, 1024-1 Arakawa, Ishigaki,
Okinawa 907-0024, Japan}

\maketitle

\begin{abstract}
The black hole binary V404 Cygni (= GS 2023+338) had an outburst on June 15, 2015 after 26 years of quiescence. 
We report the multi-color optical observation (g$^{\prime}$, R$_{\rm C}$, and I$_{\rm C}$) 
of this object at the beginning of its outburst performed by the {\it MITSuME} 50cm telescope in Akeno, Yamanashi, 
and the {\it MURIKABUSHI} 105 cm Telescope at Ishigakijima Astronomical Observatory. 
With time domain analysis of the multi-color light curves, we successfully decomposed optical variations 
into two components: a highly-variable, and a little-variable component. 
The loci of the little-variable component in the color-color diagram are consistent with that of the multi-temperature blackbody radiation, 
while those of the highly-variable component trace out a power-law spectrum with a spectral index $\alpha \sim$ 0.6--1.0. 
For the little-variable component, 
we argue that an irradiated disk with the innermost temperature higher than $\sim 2.0 \times 10^4$ K 
and the outermost temperature lower than $\sim 6.5 \times 10^3$ K is the most plausible source. 
The gradual rise trend of the light curve during our observation 
is probably due to the growing of the little-variable component. 
The observed spectral energy distribution (SED) from the optical to ultraviolet can be expressed by a model 
consisting of a power-law component and an irradiated disk component. 
\end{abstract}

\section{Introduction}
Many stellar-mass black holes are widely known to have a violent X-ray flux oscillation on various time scales 
from milliseconds to hours. 
Each time scale being related to different emission origins 
such as the non-thermal jet, the accretion disk, or the corona; 
the X-ray fluctuation we can observe is the superposition of them. 
Some models are proposed to describe the behavior of such objects, 
included in low mass X-ray binaries (LMXBs); {\it e.g.}, disk instability model, see Lasota 2001 for a review of the model. 
Its validness is confirmed for 52 X-ray binaries including 23 black hole binaries 
mainly based on X-ray flux variations (Coriat et al. 2012). 

In optical or infrared range, however, 
similar variations with X-ray's have been observed in only a few black hole binaries: 
GX 339$-$4 (Motch et al. 1983), GRS 1915+105 (Eikenberry et al. 1998), 
XTE J1118$+$480 (Uemura et al. 2000), V4641 Sgr (Uemura et al. 2002), 
and V404 Cygni (Tanaka et al. 2016, Mart\'i et al. 2016, Gandhi et al. 2016, Kimura et al. 2016). 
For GX 339$-$4 and XTE J1118$+$480, 
the optical flickering is dozens of per cent of its flux on time-scales of seconds, 
and the scenario of cyclo-synchrotron emission associated with various magnetic flare scales 
was proposed by Motch et al. (1983) and Merloni et al. (2000), respectively. 
%based on an existence of QPO ($\sim 0.1$ Hz) and spectral analysis. 
On the other hand, the optical variation in V4641 Sgr contains 
large amplitude variations ($\sim 1$ mag, corresponding to 150 per cent of its flux) 
on a time-scale of 0.01--0.10 day along with small flickering ($\sim 0.2$ mag, corresponding to 20 per cent of its flux) 
on time-scale of 100--200 s. 
%Its power spectrum was characterized by broken power-law model with break point at 240$\pm$70 sec. 
%without any QPO features. 
%Although Uemura et al. (2004) suggested that the nature of the optical variation 
%is non-thermal emission at $\sim10^4$ Schwarzschild radius, 
%the nature of the mechanism is still unclear due to dramatically rapid darkening. 

V404 Cygni, also known as GS 2023$+$338, is a black hole binary (BHB), 
originally discovered during the 1989 outburst by the {\it GINGA} satellite (Kitamoto et al. 1989, Makino et al. 1989). 
Since the discovery of V404 Cygni, 
%which was turned out to be an one of the brightest BHB in optical band in quiescence state, 
intensive photometric and spectroscopic follow-up 
have been performed and yielded well-determined physical parameters of its system; 
the central black hole mass $M_{\rm BH} = 9.0 ^{+0.2}_{-0.6} M_{\rm \odot}$ 
with a companion mass $M_2 = 0.7 ^{+0.3}_{-0.2} M_{\rm \odot}$ (Khargharia et al. 2010), 
a parallax distance 2.39$\pm 0.14$ kpc (Miller-Jones et al. 2009), 
the inclination angle $i = 67^{\circ} {}^{+3}_{-1}$ (Shahbaz et al. 1994), 
and an orbital period of 6.5 d (Casares et al. 1992). 

On 2015 June 15, 2015 (MJD 57188), V404 Cygni had an outburst after 26 years of quiescence. 
The burst was detected and reported by {\it Swift}/Burst Alert Telescope (BAT) 
and consecutively by the Monitor of All-sky Image ({\it MAXI}) instrument 
(Barthelmy et al. 2015 and Negoro et al. 2015, respectively). 
After the notifications, intensive observations with from radio to gamma-ray is performed all around the world. 

In contrast to typical blackhole transients having 
a fast rise and exponential decay profile such as A0620-00 
within one X-ray flare (sometimes together with an optical correlated flare), 
V404 Cygni showed multiple X-ray flaring activity during the outburst (Tanaka and Shibazaki. 1996, Ferrigno et al. 2015). 
For its spectral characteristics, 
Natalucci et al. (2015), Rodriguez et al. (2015), and Roques et al. (2015) 
classified the spectral states into ``off-flare'' and ``on flare'', 
and they concluded that the origin of the flare component was Comptonization. 
Based on {\it INTEGRAL} observations, the X-ray variability is originated in 
accelerations of electrons and also the variable absorption by absorbers in the line of sight. 
{\it Chandra} observations, 
revealed the detection of emission lines in the spectrum indicating disk wind emission, 
supporting the existence of such matter (King et al. 2015). 
Kimura et al. (2016), however, pointed out that the X-ray spectrum does not show 
a noticeable rise in column density when the X-ray flux sharply dropped. 
They concluded that absorption cannot be the primary cause of the time variation in their data set. 
Radhika et al. (2016) analyzed {\it Swift}/XRT and BAT observations for the 2015 outburst of V404 Cygni. 
They found that the X-ray spectra exhibited the hard, intermediate, and soft spectral state, 
and hardness intensity diagram do not show a ``q-shape'' usually observed in black hole transients 
(Fender and Belloni 2004). 
They suggested that spectral change is associated with coronal ejection.  

For optical range, 
the nature of the variation is still under discussion 
in spite of the well determined physical parameters that 
would enable us to investigate the physics around the black hole. 
The scenario that the nature of the optical emission is mainly the outer disk 
irradiated by X-ray emission generated in the inner disk, 
is proposed by Kimura et al. (2016) 
based upon repetitive patterns in the optical variation and spectral energy distribution (SED) analysis 
with a quite extensive data set of multi-color optical photometric data, 
and supported by Mu$\tilde{\rm n}$oz-Darias (2016). 

On the other hand, the scenario that relativistic non-thermal emission component with 
the accretion disk spectrum is proposed by Marti et al. (2016) and Tanaka et al. (2016). 
Mart\'i et al. (2016) tentatively interpreted the origin of the optical emission 
by the relativistic plasmons along collimated jets as a result of accretion disk instabilities. 
Tanaka et al. (2016) had further investigated the origin of the variation based upon 
modeling of broad-band SED from radio to gamma-ray bands. 
Gandhi et al. (2016) reported sub-second optical flaring activity in the optical band on MJD 57199 
(corresponding to the peak of the outburst) and explained its origin as optically thin synchrotron emission from a jet. 
Lipunov et al. (2016) and Shahbaz et al. (2016) reported the detection of the change of the polarization ($\sim$ 1--5\%)
correlating with optical variability and concluded that the variation of the polarization is yielded from relativistic jet. 
All of them suggested the existence of either a disc or the other component for the optical emission 
together with the jet. 

%To make it easy to investigate the physics relating to the flux variation quantitatively, 
%the decomposition or the categorization of the variation would give fundamental information ({\it e.g.} Belloni et al. 2000). 
In this paper, we present the optical data obtained using a MITSuME instrument 
and results acquired through spectral-timing analysis. 
Our aim is to derive information about the nature of the complex optical variation.

\section{Observations and Data Reductions}
%\subsection{Optical Bands}
A series of intensive optical observations of V404 Cyg were performed 
with three color imaging system developed for MITSuME project 
(Kotani et al. 2005, Yatsu et al. 2007, Shimokawabe et al. 2008) 
on the {\it MITSuME} 50cm telescope in Akeno and the {\it MURIKABUSHI} 105cm telescope 
during the period MJD 57192 -- 57207. 
The system allows us to take simultaneous images 
in g$^{\prime}$, R$_{\rm C}$, and I$_{\rm C}$-bands 
by employing dichroic mirrors. 
The effective wavelengths of the filters are 4770, 6492, and 8020 \AA, respectively (Fukugita et al. 1996). 
{\it MITSuME} telescope employs three Apogee Alta U-6 camera, KAF-1001E CCD 
with $1024\times1024$ active area pixels, and the pixel size is 1.6 arcsec pixel$^{-1}$ at the focal plane. 
It is designed to have a field of view (FOV) of 27.3 arcmin. 
On the other hand, {\it MURIKABUSHI} telescope covers a FOV of 12.3 arcmin, 
and then has a pixel resolution of 0.72 arcsec pixel$^{-1}$. 

The obtained raw data were preprocessed in a standard manner; 
subtracting dark and bias, and then dividing by a flat frame. 
The pixel coordinates are calibrated into celestial coordinates using WCSTools (Mink 1997). 
After the primary treatment, we performed aperture photometry to estimate the magnitude of this object 
by comparing with five local reference stars using IRAF tasks. 
In order to improve an accuracy in the magnitude estimation of V404 Cyg, 
the magnitudes of reference stars are calibrated by eleven photometric standard stars 
selected from Landolt (1992) and Smith et al. (2002).
Information of the reference stars and the standard stars are summarized in Table \ref{tab:ref}. 

In this paper, we focus on the flux variation of V404 Cyg in MJD 57192 -- 57194, 
because during this period we could obtain a higher quality photometric data 
thanks to good sky condition compared to later observations. 
We used variable exposures of 5 -- 60 sec, chosen to give reasonable count rates for this object 
in the three bands at different flux levels 
(g$^{\prime} \lesssim 15.5$ mag, R$_{\rm C} \lesssim 13.5$ mag, and I$_{\rm C} \lesssim 12.5$ mag). 
A log of the observations is given in Table \ref{tab:log}. 
It should be note that there is a line-of-sight contaminating star at 1.5 arcsec north of V404 Cyg 
and we could not separate it from V404 Cyg. 
However, the magnitude of the contamination star is g$^{\prime} \sim 19.9$ mag (Shahbaz et al. 2003), 
$\sim 60$ times fainter than V404 Cygni at its faintest flux level in this period. 
We therefore ignored the contamination in our photometry procedures. 
The time series data of g$^{\prime}$, R$_{\mathrm C}$ and I$_{\mathrm C}$-band photometry 
will be available via Centre de Donn{\'e}es astronomiques de Strasbourg (CDS, Strasbourg
astronomical Data Center
\footnote{\url{http://cds.u-strasbg.fr/}}).

\section{The Light Curve of V404 Cyg}\label{sec:lc}
%%%%%%%%%%%%%%%%%%%%
\begin{figure}
\includegraphics[width=8.75cm]{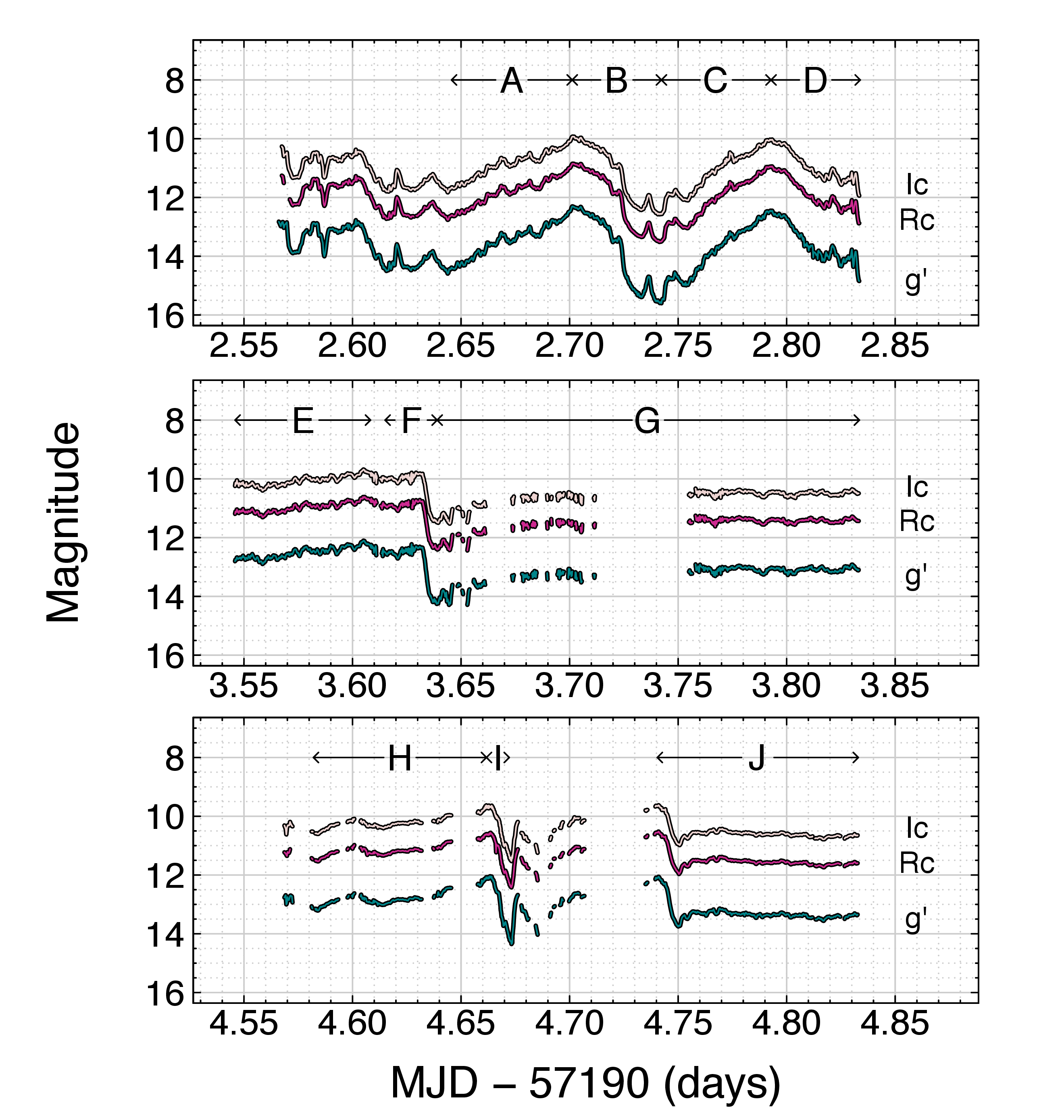}
\caption{Observed g$^{\prime}$, R$_{\rm C}$, and I$_{\rm C}$ light curves of V404 Cyg 
	in MJD 57192--57194. }
\label{fig:lc}
\end{figure}
%%%%%%%%%%%%%%%%%%%%
Fig.\ref{fig:lc} shows the optical (g$^{\prime}$, R$_{\rm C}$, and I$_{\rm C}$-band) light curves 
during MJD 57192--57194, corresponding to 4--6 days after the detection of X-ray outburst 
on MJD 57188. 
Although we continuously observed V404 Cyg in the optical bands through the night in these observational days, 
on around MJD 57193.30 or 57194.72, for example, the observation was partially intercepted by clouds 
resulting in loss of data. 

MJD 57192--57194 being categorized as ``gradually rising'' phase toward to ``plateau'' phase (Kimura et al. 2016), 
we can see that averaged brightness in each band on MJD 57194 
(13.2 mag, 11.4 mag, and 10.5 mag in g$^{\prime}$, R$_{\rm C}$, and I$_{\rm C}$-band, respectively) 
is brighter about 0.6 mag than that on MJD 57192 
(13.7 mag, 12.0 mag, and 11.1 mag in each band, respectively). 
The amplitudes of flux variations was largest on MJD 57192 reaching $\sim 3$ mag 
on a timescale shorter than an hour. 
Variation on MJD 57192 seems to be composed by two distinct fluctuations; 
big-slow swings and small-fast wiggles. 
On the other hand, in MJD 57193 and 57194, 
fast drops in flux rather than big-slow swings were seen, while small wiggles remained. 
Throughout the observations, 
variations with large amplitudes always have longer timescales in rising parts than in the falling parts. 
For all variations, the fluxes in the optical three bands were always well correlated. 

\section{Screening of Optical Photometric Data}\label{sec:scr}
In order to improve accuracy of the analysis, 
we had screened the optical data using the measurements of magnitudes of two reference stars. 
First, we discarded low-quality data with $\sqrt{ \sigma^2_{i} + \sigma^2_{j} } \ge 0.25$, 
where $\sigma_i$ or $\sigma_j$ is the statistical error in calculated magnitude of reference star $i$ or $j$. 
And next we selected the data whose 99.99\% confidence interval (for the normal distribution) of 
the difference in magnitudes of reference star $i$ and $j$, $\Delta M_{ij}$, 
including the average of the deference in two reference stars 
over the all data $\overline{\Delta M_{ij}}$ with 
$|\Delta M_{ij} - \overline{\Delta M_{ij}}| \le 3.89 \sqrt{ \sigma^2_{i} + \sigma^2_{j} }$. 

The resulting typical statistical uncertainty in the magnitudes of the reference stars 
were $\sim 0.07$, $\sim 0.03$, and $\sim 0.04$ {\bf mag} 
in g$^{\prime}$, R$_{\rm C}$, and I$_{\rm C}$-bands in {\it MITSuME}/Akeno, 
and $\sim 0.01$ mag in the all optical bands in the {\it MURIKABUSHI} Telescope. 
The error of $\Delta M_{ij}$ ($= \sqrt{\sigma^2_{i} + \sigma^2_{j}} $) larger than 0.25 was therefore 
outlier, mainly due to bad sky condition. 
We iterated the latter treatment until $\overline{\Delta M_{ij}}$ converged. 
The data selections are performed for all combinations of reference stars (${}_5 C _2 = 10$ possible outcomes), 
and we picked up the data meeting the requirement for the entire twenty combinations. 
Eventually, about 35\% data are discarded by this treatment.

\section{Analysis and results}
\subsection{Color-color Diagram}\label{sec:ccd}
%%%%%%%%%%%%%%%%%%%%  
\begin{figure}
\includegraphics[width=8.75cm]{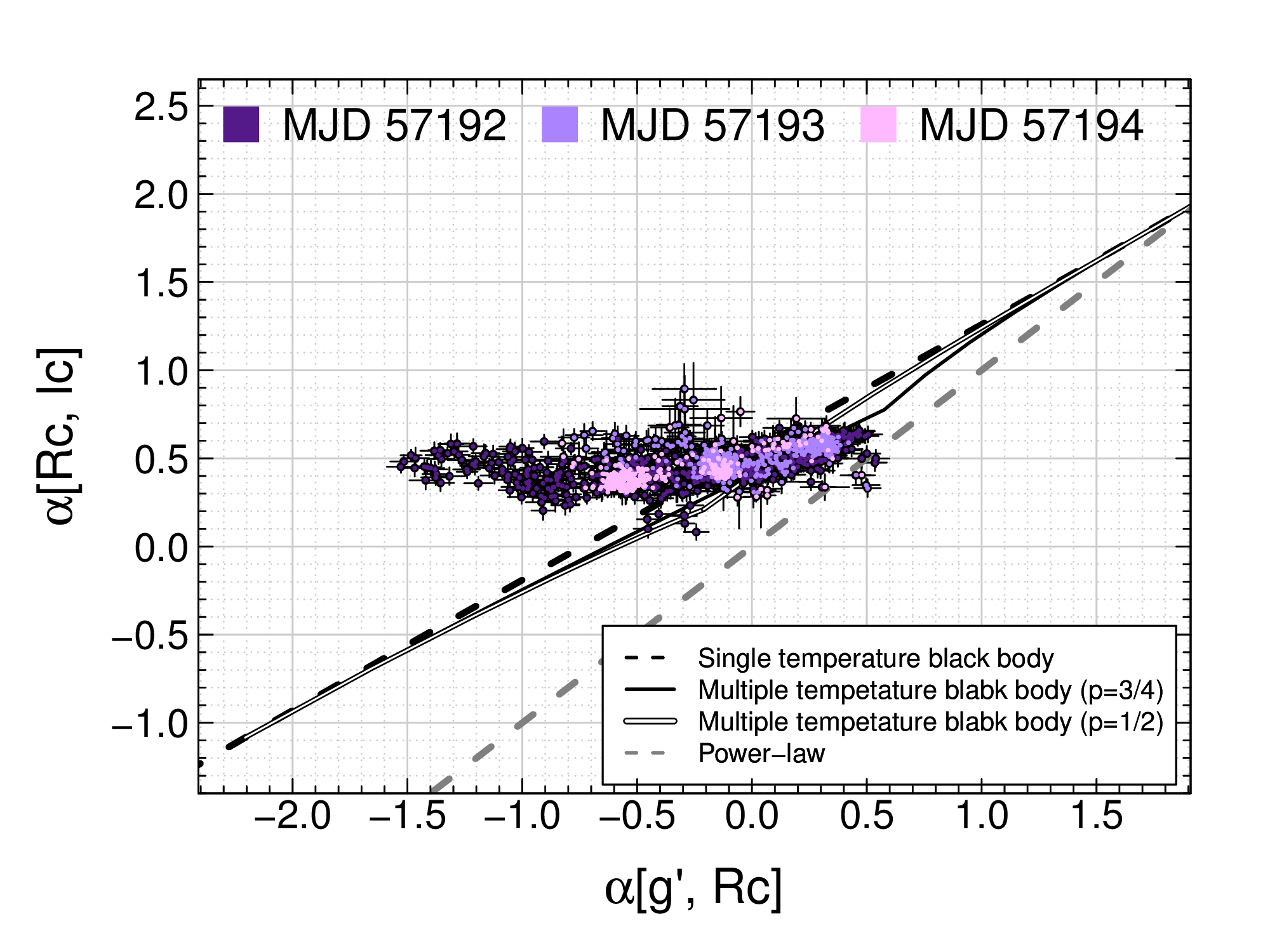}
\caption{Color-color diagram for the data in Fig.\ref{fig:lc} 
	with the spectral indices between g$^{\prime}$-band and R$_{\rm C}$-band, 
	and R$_{\rm C}$-band and I$_{\rm C}$-band. 
	Magnitude is converted into energy flux density $F_{\nu}$ [erg/s/cm$^2$/Hz] 
	using A$_{\rm V} = 3.3$ and R$_{\rm V} = 2.8$. The details are written in \S\ref{sec:ccd}. }
\label{fig:col}
\end{figure}
%%%%%%%%%%%%%%%%%%%%
Color-color diagram is helpful to categorize the observed flux variation and 
to test the applicability of assumed emission models.
Belloni et al. (2000), for example, have classified X-ray variation in GRS 1915$+$105 into twelve separate classes 
and proposed the emission processes of these variations through color-color diagram analysis. 
We therefore calculated the spectral indices 
between g$^{\prime}$-band and R$_{\rm C}$-band ($\alpha_{gr}$), 
and R$_{\rm C}$-band and I$_{\rm C}$-band ($\alpha_{ri}$) for all the data point in Fig.\ref{fig:lc} 
and presented them in a color-color diagram (Fig.\ref{fig:col}). 
When converting the observed magnitudes to intrinsic fluxes $F_{\nu}$,
we adopted $A_V = 3.3$ and $R_V= A_V / E(B-V) = 2.8$ 
among the possible values, 
$A_V = $ 3.0 -- 3.6 (Shahbaz et al. 2003, Hynes et al. 2009, and Itoh et al. 2016) 
with $R_V <$ 3.0 (itoh et al. 2016), 
under the constrain of $E(B-V)\sim$ 1.2 (itoh et al. 2016; Rahoui et al. 2017). 
Then the dereddening law described in O'Donnell (1994) was applied 
for the central frequencies for the three bands: 
$3.74\times10^{14}$ Hz (I$_{\mathrm C}$-band), 
$4.62\times10^{14}$ Hz (R$_{\mathrm C}$-band), and 
$6.28\times10^{14}$ Hz (g$^{\prime}$-band).
The result is drawn by grey points in Fig.\ref{fig:col}. 
The grey dashed line and black dashed line trace loci for power-law spectra with different spectral indices 
and single temperature black body with different temperatures, respectively,  
whereas the black line and black-edged white line represents 
those for standard disk blackbody and irradiated disk blackbody
(p=3/4: Shakura \& Sunyaev 1973, and p=1/2: King \& Ritter 1998) 
for different inner/outer-most temperatures under the condition that the disk temperature cannot  
across the critical temperature $\sim 5000 K$ where the disk would be unstable 
(King \& Retter 1998, ${\rm {\dot Z}}$ycki et al. 1999), respectively. 
Here, $p$ is the gradient index of the surface temperature of the accretion disk ($T$) 
when expressed as a function of the radial distance from the center ($r$); {\it i.e.} $T \propto r^{-p}$
({\it e.g.} Mineshige et al. 1994). 
If the optical emission has a power-law spectrum 
and its variation is attributed to only the change of its normalization, 
the data points will stay at the same place on the grey dashed line. 
On the other hand, if the power-law spectrum changes its spectral index, 
data point move along the grey dashed line. 
Similarly, points for a blackbody or disk emission model follow the corresponding line 
if its characteristic temperature changes.

Apparently, the points in the color-color diagram does not trace any of the model lines 
of the single temperature black body, the multi-temperature black body, or power-law spectra. 
It indicates that 
the optical variation cannot be attributed to a single emission component, 
expressed by single-temperature or multi-temperature blackbody, or power-law model. 

\subsection{Flux-flux Plot} \label{sec:ffp}
%%%%%%%%%%%%%%%%%%%%
\begin{figure}
\includegraphics[width=8.75cm]{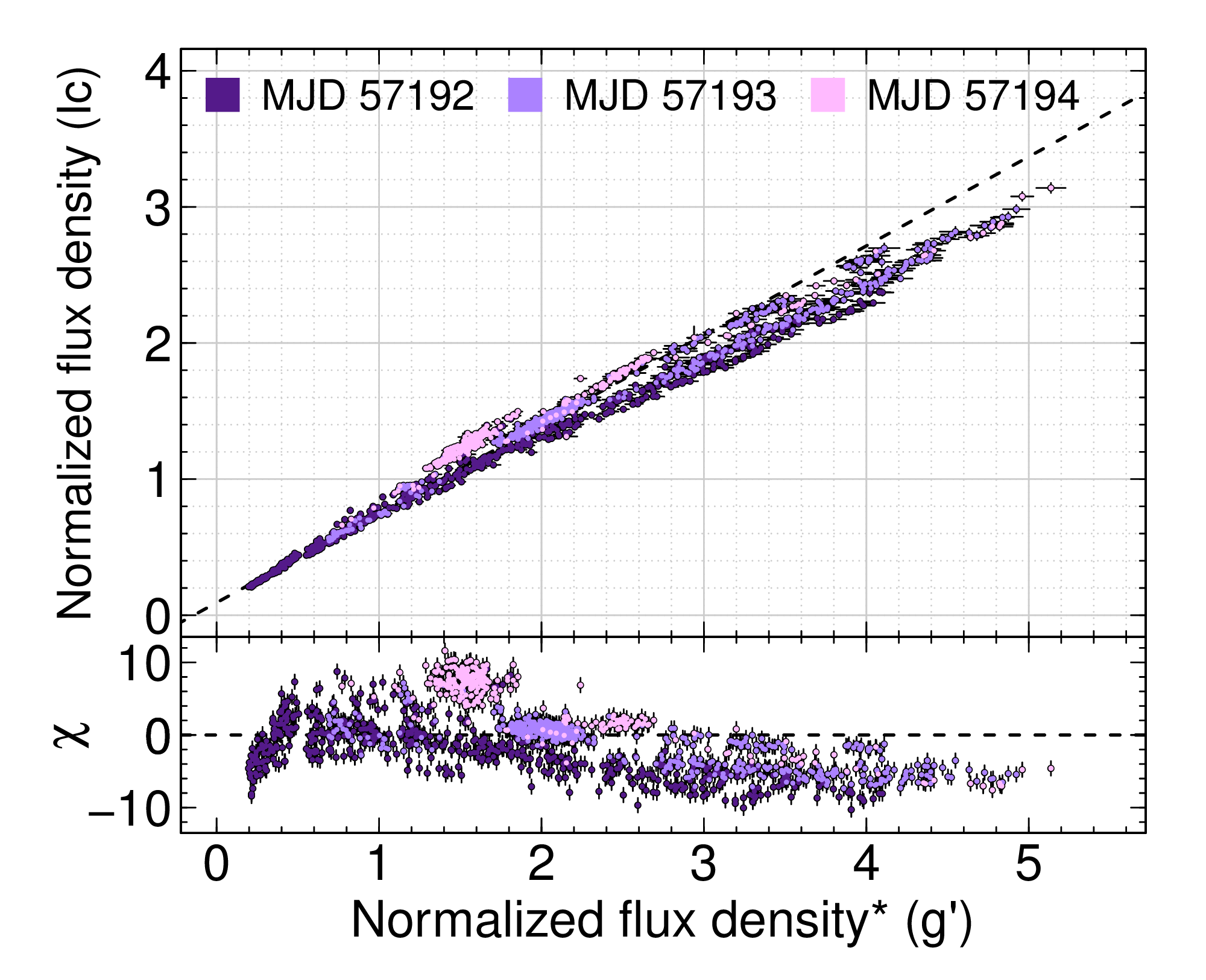}
\caption{Flux-flux plot with the fluxes between g$^{\prime}$-band and I$_{\rm C}$-band. 
	The details are written in \S\ref{sec:ffp}. }
	\footnotesize{${}^{\ast}$In units of $10^{23}$ erg/cm$^2$/s/Hz assuming tentative value 
	A$_{\rm V} = 4.0$ and R$_{\rm V} = 3.1$. Note that the absolute value of the flux 
	does not influence our discussion and result.}
\label{fig:ffp}
\end{figure}
%%%%%%%%%%%%%%%%%%%%
%%%%%%%%%%%%%%%%%%%%
\begin{figure*}[t]
\begin{center}
\includegraphics[width=17.0cm]{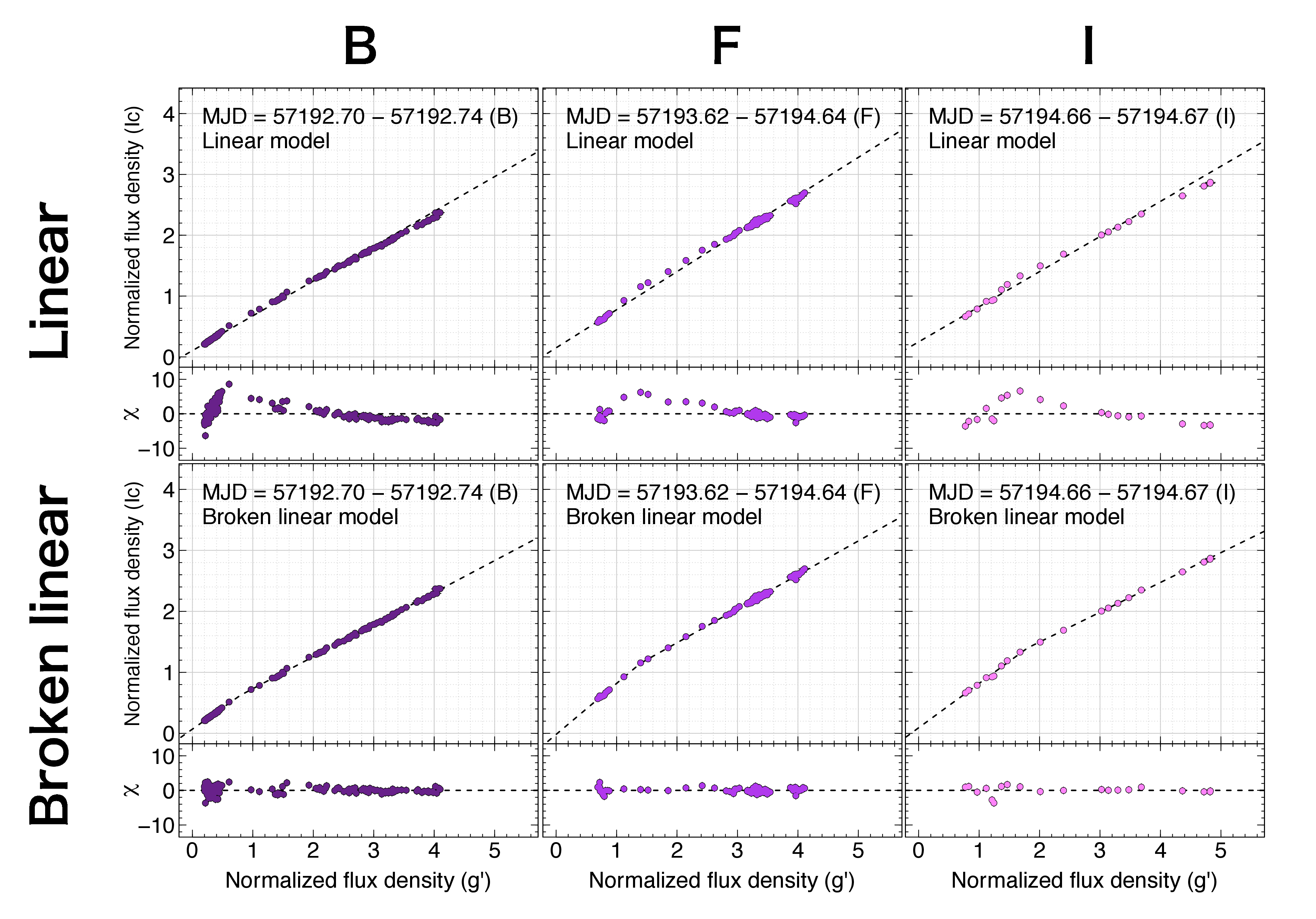}
\end{center}
\caption{The typical flux-flux plots with the fluxes between g$^{\prime}$-band and I$_{\rm C}$-band. 
	The upper panels are those fitted with a liner model, 
	while the lower panels are fitted with a broken liner model. 
	The details are written in \S\ref{sec:mod}.}
\label{fig:ffp_fit}
\end{figure*}
%%%%%%%%%%%%%%%%%%%%
Using the color-color diagram, we showed that 
the optical variation of V404 Cyg is not attributed to changes of a parameter of 
a single emission mechanism. 
In this section, we further investigate the properties of the optical behavior phenomenologically 
on the flux-flux plot. 

Fig.\ref{fig:ffp} shows the flux-flux plot constructed by the g$^{\prime}$ and I$_{\rm C}$-band photometric data 
from MJD 57192 to 57194. %, dereddened assuming $A_V = 4.0$. 
Gray scale of each point indicates the date of the observation, as shown in the right bar in MJD, and 
the dashed line on the upper panel shows the linear fit to all the data, 
where fitting was performed to minimize the normalized residual defined as follows 
\begin{equation}
\chi^2 = \sum_{i} \frac{[y_i - f(x_i)]^2}{\sigma_{y_i}^2 + [{\rm d}f(x)/{\rm d}x]_{x_i}^2 \sigma_{x_i}^2 } \ . 
\end{equation}
Here ($x_i, y_i$) is the flux data ($x_i$: $g^{\prime}$, $y_i$: I$_{\rm C}$), 
($\sigma_{x_i}, \sigma_{y_i}$) is the corresponding uncertainties, and $f(x)$ is the function to be fitted. 
The lower panel shows the normalized residual expressed by
\begin{equation}
\chi_i = \frac{y_i - f(x_i)}{\sqrt{\sigma_{y_i}^2 + [{\rm d}f(x)/{\rm d}x]_{x_i}^2 \sigma_{x_i}^2 }} \ . 
\end{equation}
Interestingly, in the normalized residual plot, we find a general convex shape and systematic deviations 
that evolve upward ({\it i.e.} the spectrum gradually reddened as time proceeded) with time. 
Similar break features were reported in the X-ray flux-flux plots of Seyfert galaxy 
NGC 4051 (Ponti et al., 2006) and NGC 3227 (Noda et al., 2014). 
This behavior in the flux-flux plot suggests that the observed optical variation arises from at least two components 
that are dominant below and above the break. 
Our result suggests the existence of a multiple engines in optical flux variation of V404 Cygni. 
\subsection{Modeling the convex shape for the Flux-flux Plot}\label{sec:mod}
In this section, we test two functional forms to express the convex shape on the flux-flux plot: 
broken-linear function and power-law function. 

When a point ($F(\nu_1, t)$, $F(\nu_2, t)$) lies on a linear model ($F(\nu_2, t) = k_{\rm lin} F(\nu_1, t) + C$), 
%on the flux-flux plot is well expressed with the linear model ($y=k_{\rm lin}x+C$), 
the local spectral index of a variable component $\alpha$ at the two frequencies ($\nu_1$, $\nu_2$) 
can be directly derived from $k_{\rm lin}$. 
%the $k_{\rm lin}$ can directly derive from the local spectral index 
%at the two frequencies ($\nu_1$, $\nu_2$). 
%connected %on one-to-one correspondence 
%with the spectral index between different frequencies used for the abscissa and the ordinate. 
%If we think that the frequencies for abscissa and  is expressed by $\nu_1$ 
%and that to the ordinate is by $\nu_2$, 
%the optical fluxes in different energy $F(\nu_1,t)$ and $F(\nu_2,t)$ can be locally expressed 
%with spectral index $\alpha$ as follows: 
Using the local spectral index $\alpha$, $F(\nu_1)$ and $F(\nu_2)$ can be expressed by
\begin{equation}
\begin{array}{ll}
F(\nu_1,t) = A(t) \nu_1^{\alpha} + C(\nu_1) \\
F(\nu_2,t) = A(t) \nu_2^{\alpha} + C(\nu_2), 
\end{array} 
\end{equation} \label{eq:fluxes}
where $A(t)$ is normalization of the variable component, 
and $C(\nu)$ is the spectrum of stable component. 
% as a function of frequency. 
Eliminating $A(t)$, equations (\ref{eq:fluxes}) are reduced to 
%, we get the relation between $F(\nu_1,t)$ and $F(\nu_2,t)$ is given by 
\begin{equation}
F(\nu_2,t) = \left( \frac{\nu_2}{\nu_1} \right)^{\alpha} F(\nu_1, t) + \left[ - \left( \frac{\nu_2}{\nu_1} \right)^{\alpha} C(\nu_1) + C(\nu_2) \right]. 
\end{equation}  \label{eq:liner}
One see that the slope of the linear model on the flux-flux plot is $(\nu_2/\nu_1)^{\alpha}$. 
%When the variation component does not change the spectral index but change the normalization, 
% flux variation is not attributable to change of the spectral index but the normalization, therefore, 
%data points on the flux-flux plot trace a linear model. 
%the locus on the flux-flux plot is to trace a straight line. 

In contrast, if the variable component changes its spectral index, 
%spectral index (and also normalization) is varying, 
the relationship between $F(\nu_1,t)$ and $F(\nu_2,t)$ will not be linear. 
In particular, if the spectral variability is explained by power-law pivoting 
({\it e.g.} Taylor et al. 2003, Uttley \& McHardy 2005) at a frequency $\nu_p$, 
the data on the flux-flux plot should follow 
$F(\nu_2,t) - C(\nu_2) = F_{\mathrm p}^{1- \beta }[F(\nu_1,t)- C(\nu_1)]^{\beta}$, 
by reducing variable spectral index $\alpha(t)$ from 
\begin{equation}
\begin{array}{ll}
F(\nu_1,t) = F_{\mathrm p} \left( \nu_1/\nu_{\mathrm p}\right)^{\alpha(t)} + C(\nu_1) \\
F(\nu_2,t) = F_{\mathrm p} \left( \nu_2/\nu_{\mathrm p}\right)^{\alpha(t)}+ C(\nu_2) , 
\end{array} 
\end{equation} \label{eq:fluxes}
where $F_{\mathrm p}$ is the constant flux at $\nu_{\mathrm p}$ and $\beta = \log(\nu_2/\nu_p)/\log(\nu_1/\nu_p)$. 
% are constants represent by $\nu_1$ and $\nu_2$, and $\nu_p$ 
%related to $\nu_1$ and $\nu_2$  

As noted in \S\ref{sec:ffp}, the data distributions in the flux-flux plot apparently do not trace a linear model. 
We thus tested two convex functional forms to fit the flux-flux plot: 
broken-linear model (two linear models connected at a break) and power-law function.
%The normalized residual showing a convex shape, 
%we examine the applicability for fit to flux-flux plot of two functional forms: independent straight lines 
%for above and below the break point, and power-law function. 
%Broken-linear model, therefore, might be suitable for expressing the data distribution. 

%To easily interpretation of fitting results, 
Because the data in the flux-flux plot shows systematical time-dependent evolution, 
we divided the data to ten sections (A--J in Fig.\ref{fig:lc}) by their rising and declining trends. 
%light curve into a large flux enhancement or a large flux decline 
%as expressed by \UTF{00E2}\UTF{0080}\UTF{009C}A\UTF{00E2}\UTF{0080}\UTF{009D} to \UTF{00E2}\UTF{0080}\UTF{009C}J\UTF{00E2}\UTF{0080}\UTF{009D} in Fig.\ref{fig:lc} 
Note that Fig.\ref{fig:lc} shows data points before the screening described in \S4 is applied. 
Then we performed fits with broken-linear model represented to 
%We first fitted the data with a following function; 
\begin{equation}
y = \left\{ \begin{array}{ll}
	k_1 x + C & (x < F_{\rm br}) \\
	k_2 x + (k_1 - k_2) F_{\rm br} + C & (x \ge F_{\rm br}), 
	\end{array} \right. 
\end{equation}
where $k_1$ and $k_2$ is the slope of two linear functions, and $C$ is the intercept of the linear function 
for the lower flux. 
%before the break point $x = F_{\rm br}$. 
All of these four parameters are free. 
Typical examples of the fitting are shown in Fig.\ref{fig:ffp_fit}. 
The upper and lower panels are the results of fitting with the linear model and the broken-linear model respectively. 
The three columns are corresponding to the time portions 
which contain drastic flux declines: MJD = 57192.70 -- 57192.74 (B), 57193.61 -- 57193.64 (F), 
and 57194.66 -- 57194.67 (I). 
The linear model fits give $\chi^2/{\rm d.o.f} = 843/140$, $153/70$, and $130/19$ for each time periods, 
whereas the broken-linear model fits give significantly improved values $\chi^2/{\rm d.o.f} = 154/138$, $34/68$, and $30/17$. 

We next tested the power-law model: $y= k_{\rm pow} (x - C_1)^{\beta} + C_2$, 
where $k_{\rm pow}$, $\beta$, $C_1$, and $C_2$ are free parameters. 
We found power-law model give consistently worse fit than the broken-linear model for each portion: 
$\chi^2/{\rm d.o.f} = 180/138$, $55/68$, and $48/17$. 
In addition, $C_1$ and $C_2$ changed significantly from a section to another; 
though they represent the component constant during each section 
and therefore expected not to change dramatically between different section. 
%the $C_1$ and $C_2$, which are expected to be a stable component in $x$ and $y$, 
%changes significantly and downed to negative values in some cases. 
We therefore adopt a broken-linear model 
to express the behavior of the data on the flux-flux plot in this paper. 

\subsection{Decomposing the Optical Variations}\label{sec:dec}
%%%%%%%%%%%%%%%%%%%%
\begin{figure}
\includegraphics[width=8.5cm]{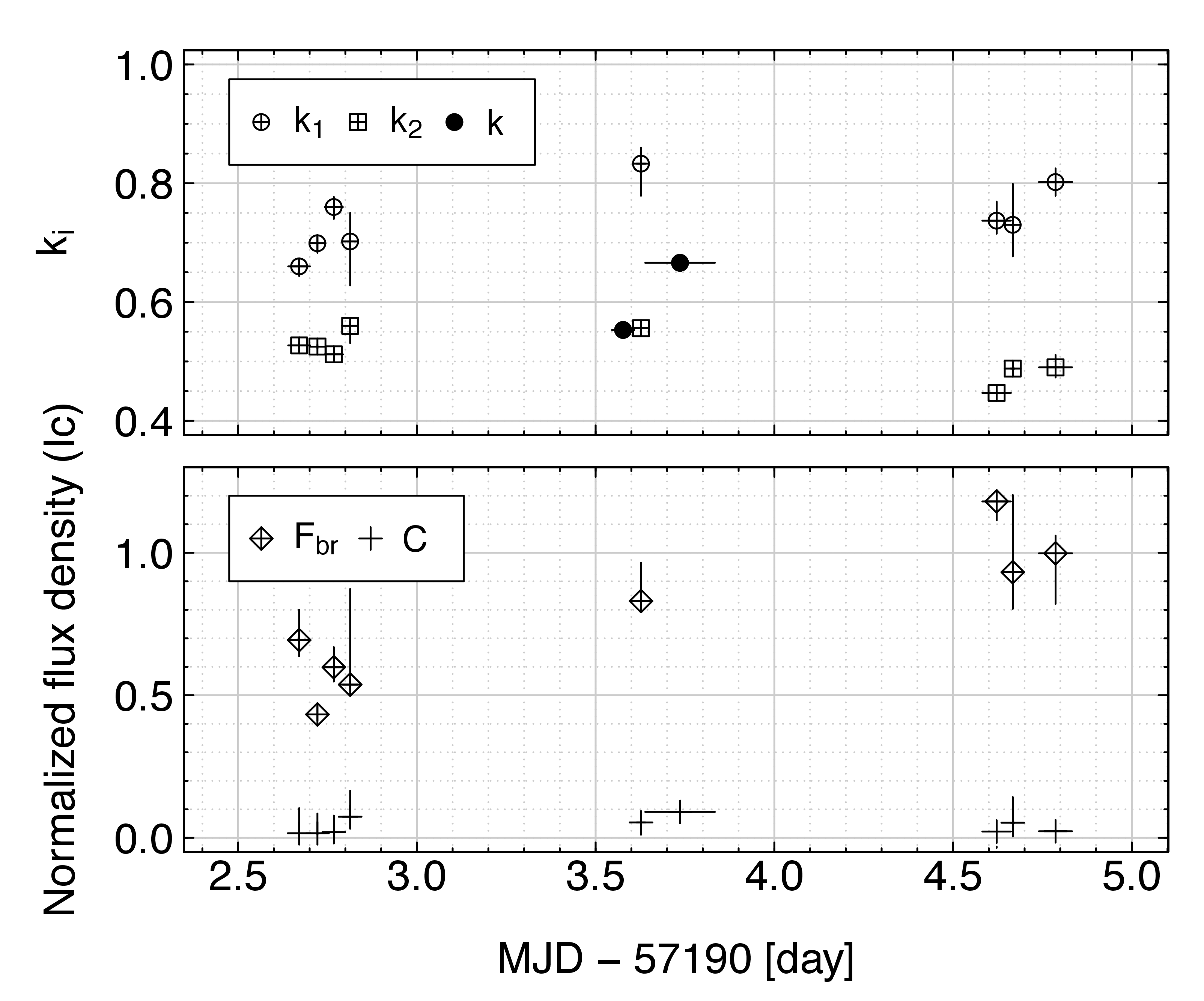}
\caption{The parameters derived by broken-liner model fitting to the flux-flux plots of g$^{\prime}$-band and I$_{\rm C}$-band. }
\label{fig:par}
\end{figure}
%%%%%%%%%%%%%%%%%%%%
As mentioned in \S\ref{sec:mod}, the slope of the straight line on a flux-flux plot is directly leads the local spectral index 
between $\nu_1$ and $\nu_2$. 
For the optical variation in V404 Cygni, we found that the data in flux-flux plot was 
well expressed by a broken-linear model (see \S\ref{sec:mod} and Fig.\ref{fig:ffp_fit}). 
Two linear branches below and above a break imply that the existence of two variable components 
with distinct spectral indices. 

To derive slopes for two branches, we fitted a broken-linear model to data in entire periods A--J. 
It must be noted that 
(1) we used ``screened data" explained in \S\ref{sec:scr}, 
(2) we only used portions containing more than 15 data points, 
and 
(3) we used a single component linear model for the portions where a break was not prominent. 
Except for such two portions, MJD = 57193.5459--57193.6083 (E) and 57193.6390--57193.8335 (G), 
we derived two slopes with broken-linear models. 
The derived parameters are tabulated in Table \ref{tab:par_sum}. 
Those for the g$^{\prime}$ and I$_{\rm C}$ pair are plotted in Fig.\ref{fig:par}. 
The slopes of the two segments of the broken linear model ($k_1$ and $k_2$) at each portions, 
we find that all of the $k_1$ and $k_2$ are clearly separated at $\sim$0.6 (Fig.\ref{fig:par}, the upper panel). 
For the two portions E and G, thus, 
we assigned the slope $k$ to $k_1$ 
if it is larger than 0.6, and to $k_2$ otherwise. 
Additionally, we found in flux-flux plots (${\it e.g.}$ Fig.\ref{fig:ffp_fit}) that 
the dynamic range of the flux variation in the component above the break is lager  
than that below.  
For example, the variation amplitude of the g$^{\prime}$-band flux below and above the break 
are $3\times$ and $18\times$ the minimum flux respectively in period ``B'' 
(leftmost panel of Fig.\ref{fig:ffp_fit}). 
It indicates that one variable component has quite large variability and 
dominates the optical variation when the component is much brighter than another one. 
Therefore, we named the components dominated at above the break as ``highly-variable component" (HVC) 
and below that as ``little-variable component'' (LVC). 
The slope (or the flux ratio to g$^{\prime}$ band) of the LVC is varying in intra-day and has a redder color, 
whereas that of the HVC $k_2$ is relatively stable but has long-term change and a bluer color. 
\subsection{Decomposed Color-color Diagram} \label{sec:rsed}
%%%%%%%%%%%%%%%%%%%%
\begin{figure}
\includegraphics[width=8.75cm]{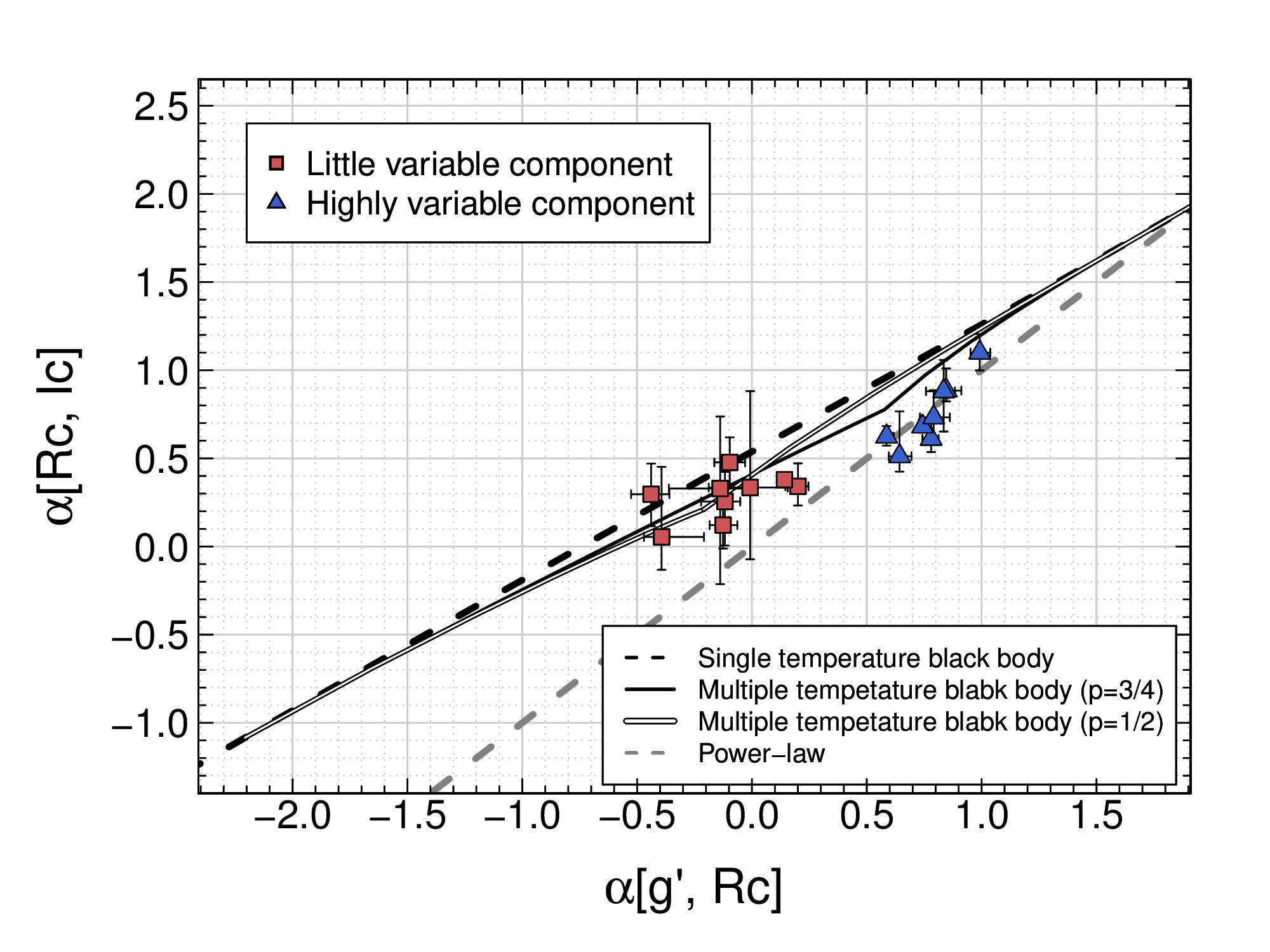}
\caption{Color-color diagram with the spectral indices between g$^{\prime}$-band and R$_{\rm C}$-band, 
	and R$_{\rm C}$-band and I$_{\rm C}$-band, after the decomposition. }
\label{fig:dec_col}
\end{figure}
%%%%%%%%%%%%%%%%%%%%
Fig.\ref{fig:dec_col} shows the color-color diagram for the decomposed components (HVC and LVC) 
constructed from the slopes of the broken-linear model.
A variety of spectral indices of power law spectra should fall on the grey dashed line in the same way as Fig.\ref{fig:col}. 
The data points below the grey dashed line, therefore, 
corresponds to convex spectra and vice versa above the line. 

Although spectral indices of the original data between g$^{\prime}$ and R$_{\rm C}$ 
reach down to $-1.5$ as shown in Fig.\ref{fig:col}, 
those after the decomposition never reach $-0.5$ as shown in Fig.\ref{fig:par}. 
It is due to the redder stable component 
(second term of the right hand side in Eq.\ref{eq:liner} and constant $C$ in Table \ref{tab:par_sum})
especially prominent in the R$_{\rm C}$-band. 

The spectral indices of the HVC lie on 
the gray dashed line indicating power-law spectra with variable its indices. 
On the other hand, those of the LVC exhibits softer spectra: 
spectral indices are consistent with those of the standard disk (p=3/4)
with the innermost temperature (T$_{\rm in}$) of typically $\sim 1.5 \times 10^4$ K and 5000 K at the outer edge (T$_{\rm out}$), 
or the irradiated disk (p=1/2) with T$_{\rm in} \gtrsim 2\times10^4$K and T$_{\rm out} \lesssim 6500$K. 

\section{Discussion}
\subsection{Optical Emission Components} \label{sec:com}
We have shown that the optical variation is decomposed to two components; 
little-variable component (LVC) and highly-variable component (HVC). 
HVC has larger amplitude ($\sim$5--10 times larger than that of LVC) and bluer color, 
and thus corresponds to ``big-slow swings" mentioned in \S\ref{sec:lc}.  
On the other hand, 
LVC which has a redder spectrum corresponds to a characteristic behavior 
mentioned in \S\ref{sec:lc} as ``small-fast wiggles" as is similar to ``heartbeat-type" oscillations 
observed by Kimura et al. (2016), 
considering its small amplitude and its dominance at the low flux phase. 

Besides these components, Gandhi et al. (2016) reported rapid optical flux variability 
which sometimes are unresolved down to a time resolution of 24 milliseconds in the outburst. 
The rapid fluctuation is only present near the peak of the outburst on MJD 57199, 
but not on MJD 57193, 57194 and 57198. 
Although we are not sure whether sub-second flares were present in our data, 
it is reasonable to assume that the contribution from the sub-second flare is negligible because,
(1) our dataset is in MJD 57192--57194, when the fast flares unlikely present, 
and (2) the variation amplitude of the sub-second flare ($\sim$0.2 mag, Gandhi et al. 2016) is adequately small 
relative to that from the HVC or LVC ($\sim$3 mag and $\sim$1 mag). 
The sub-second optical flaring activity therefore would not affects our analysis and discussion below.

\subsection{H$\alpha$ Emission Line}\label{sec:ha}
Gandhi et al. (2016) calculated the contribution of H$\alpha$ emission line to the r$^\prime$-band photometry 
by modeling the continuum spectrum under assumptions of spectral index 
ranging from +2 for a blue continuum to $-$2 for a red continuum. 
The result is that H$\alpha$ contributes about 10\% to their r$^\prime$-band flux at MJD=57199. 
If the H$\alpha$ contribution is in R$_{\rm C}$-band, the spectral index 
between R$_{\rm C}$-band and I$_{\rm C}$-band will be reduced by about 0.5, 
whereas that between g$^{\prime}$-band and R$_{\rm C}$-band increases by about 0.4. 
The data points in Fig.\ref{fig:dec_col} would then move toward lower-left region, 
and then almost all of them depart the power-law loci. 
If HVC or LVC contains the H$\alpha$ line, therefore, 
the underlining continua of these components must have unnatural concave shapes 
at about every portions of our data set. 

Additionally, Mu{\~n}oz-Darias et al. (2015) found that 
flux of the H$\alpha$ line was stable regardless of the variable continuum on MJD 57192. 
The H$\alpha$ line, therefore, 
should be contained in the stable component (columns $C$ in Table \ref{tab:par_sum}). 

\subsection{Origin of LVC}\label{sec:lvc}
The time-scale of the LVC ({\it i.e.}, ``small-fast wiggles'') about a few minutes, 
corresponding to the light crossing time of $\sim 10^6$ Schwarzschild radius, 
constrains the upper limit for the optical-emission area of the accretion disk. 
It is roughly consistent with the radii where the disk can extend out, 
$1.2 \times 10^{12}\ {\rm cm} = 4.4\times10^5$ Schwarzschild radius (Shahbaz et al. 1994)
for the mass of the black hole M$_{\rm BH} \sim 9\ M_{\odot}$. 

We found that spectral indices of LVC are consistent with those of a standard disk 
or an irradiated disk. 
However, a standard disk truncated at such a low innermost temperature ($\sim 1.5 \times 10^4$ K) seems to be 
difficult to reconcile with the detection of a disk component in the soft X-ray region (Radhika et al. 2016 and Walton et al. 2016). 
For the origin of LVC, thus, the irradiated disk is favored. 
From the observed flux, the outer radius of the disk is evaluated to be $\sim$ 3--5 $\times 10^5$ Schwarzschild radius. 
It is consistent with the size limit based on the timescale of LVC.  
The irradiated disk is therefore the most plausible source of the LVC. 

We additionally note that the gradual brightening during MJD 57192 -- 57195 
(also reported by Kimura et al. 2016) is likely due to increase of the LVC. 
In the lower panel of Fig.\ref{fig:par}, the break flux density $F_{\rm br}$ increases day by day 
while the intercept $C$ stays almost constant. 
$F_{\rm br}$ at MJD 57914 is about two times higher (corresponding to $\sim$0.75 mag lower) than that at MJD 57192. 
This means that the gradual brightening trend is mostly attributed to the increase of the average flux of LVC, 
probably because the optical emitting region gradually moved outward as X-ray emission from the central region grew stronger. 
\subsection{Origin of HVC}
The decomposed color-color diagram shows that the spectrum of HVC is following power-law model 
with a spectral index of $\sim$ 0.6--1.0. 
This component has drastic amplitude of variation ($\sim$5--10 times larger than that of LVC) within $\sim$ 15--50 minutes, 
and thus corresponds to ``big up-and-down swings'' mentioned in \S\ref{sec:lc}. 
The rising time (T$_{\rm r}$) of flux is constantly longer than the decaying time (T$_{\rm d}$), 
while T$_{\rm r}$ had changed day by day; 
on MJD = 57192, T$_{\rm r} \sim 1 $ hour is slightly longer than T$_{\rm d} \sim 30$ min, 
whereas on MJD = 57193, T$_{\rm r} > 4$ hour is apparently longer than T$_{\rm d}$. 
Though, interestingly, the spectral index of HVC is almost stable for all of our data set. 
For interpreting power-law spectrum of HVC, 
synchrotron emission from the jet is conceivable firstly. 

Mart\'i et al. (2016) discussed about the possibility that 
the optical variation in flux and color is interpreted in an scenario based on the ejection of non-thermal emitting such as jet 
from radio to optical wavelengths. 
Recently, Shahbaz et al. (2016) and Lipunov et al. (2016) reported 
variable optical linear polarization, changing by $\sim$ 1\% over a timescale of $\sim$ 0.5--1 hour, 
after subtracting the contribution from interstellar dust between V404 Cyg and the Earth. 
The time-scale of the variation is consistent with that in our ``big up-and-down swings''. 
Both of them concluded that the jet synchrotron emission is a viable candidate for the variable polarization. 
Furthermore, during this outburst V404 Cyg exhibited large variation in the radio 
(Tetarenko et al. 2015, Trushkin et al. 2015b), 
and its spectrum at MJD 57191.952 can be well described by power-law with spectral index $\sim 0.65$ 
(Trushkin et al. 2015a). 
Therefore, a jet was likely to be present. 
Since the spectral break from optically thick regime to thin 
is expected to lie at $\sim$ 1.4--4.7 $\times 10^{14}$ Hz (Shahbaz et al. 2016, Tanaka et al. 2016, and Itoh et al. 2016), 
the blue spectrum seen in HVC ($\alpha$ $\sim$ 0.5--1.0) is probably attributed to 
the transitioning spectrum from optically thick synchrotron spectrum to that of thin regime. 
The jet is therefore one of the strongest candidate of the origin of HVC. 

Another possibility is the thermal cyclo-synchrotron emission as proposed for  
XTE J1118$+$480 (Merloni et al. 2000 and Kanbach et al. 2001) and V4641 Sgr (Uemura et al. 2002). 
The time-scale of the optical fluctuation in V4641 Sgr is similar to V404 Cygni (shorter than $\sim 1$ hour),  
and moreover the self absorbed regime of the cyclo-synchrotron spectrum can produce the spectral index harder than $\sim$0.6. 
The cyclo-synchroron emission is also a possible mechanism of the power-law component (HVC). 
\subsection{Origin of the stable component}
As mentioned in \S\ref{sec:rsed}, the stable component, 
namely a component not varying much in the time window of a flux-flux plot, 
is an underlying variable component prominently in the R$_{\rm C}$-band (Tab.\ref{tab:par_sum}, denoted as $C$). 
Its time-scale of the variation, therefore, must be longer than $\sim$1 hour. 
Although we cannot determine the spectral shape, 
it is likely that H$\alpha$ emission is contained in this component, 
because H$\alpha$ strength seems to be more stable and almost continuum independent (Mu{\~n}oz-Darias et al. 2015), 
and the flux of this component is prominent in the R$_{\rm C}$-band which is including the wavelength of the line. 
These pieces of evidence, 
namely, the variation time scale longer than $\sim$ 1 hour and the possibility of the line contribution in R$_{\rm C}$-band, 
suggest that the stable component is a disk or companion wind origin ({\it e.g.}, Radhika et al. 2016, Rahoui et al. 2017). 
Its flux contribution is typically larger than $\sim$5--10\% (estimated by $C/F_{\rm br}$) in MJD 57192, 
and then gradually decrease toward $\sim$2--5 \% as $F_{\rm br}$ grow up (mentioned in \S\ref{sec:lvc}). 

\subsection{SED from optical to UV}\label{sec:sed}
%%%%%%%%%%%%%%%%%%%%
\begin{figure}
\includegraphics[width=8.5cm]{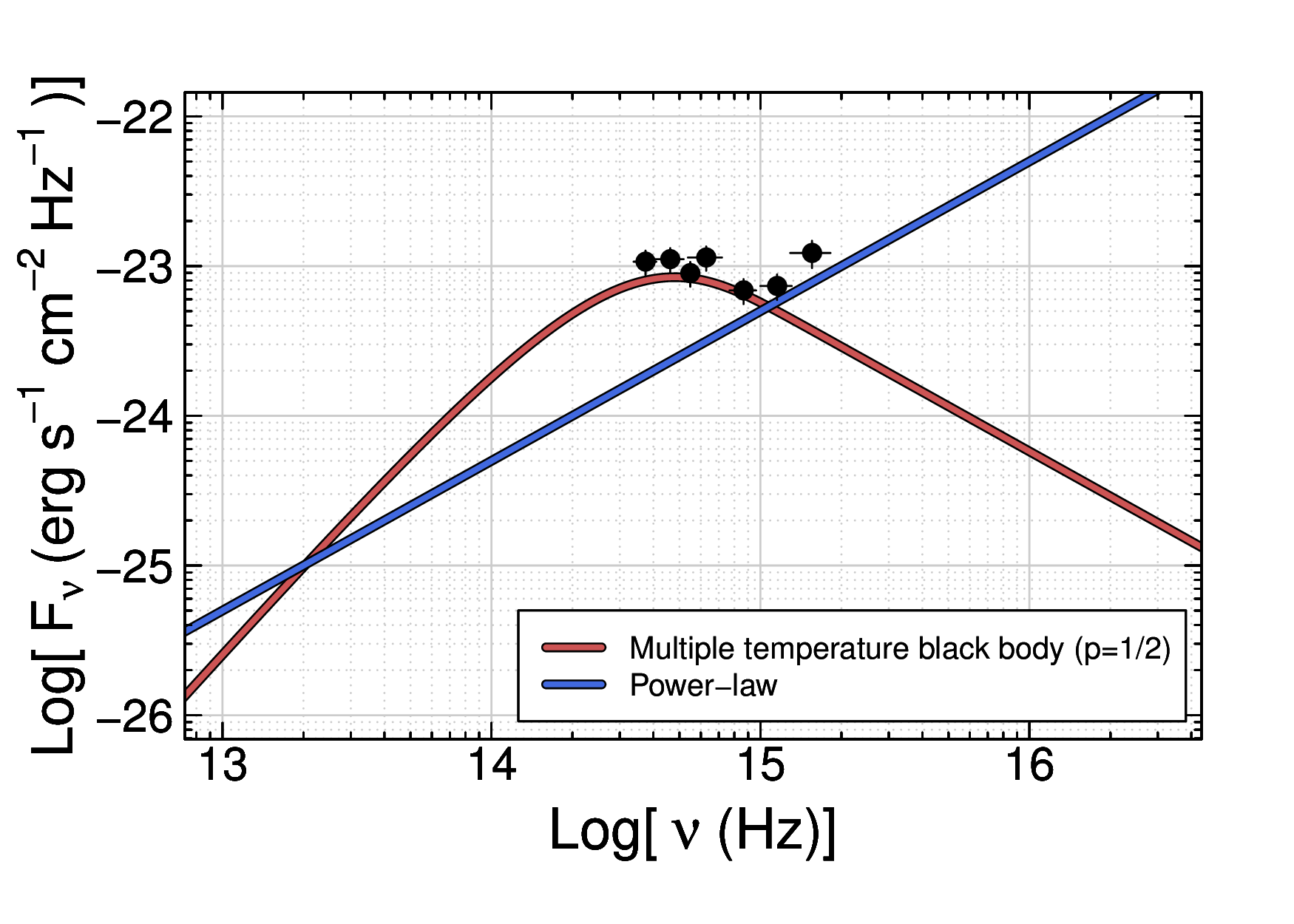}
\caption{SED from optical to UV with an irradiated disk spectrum (solid line) and power-law spectrum (dashed line) in MJD $\sim$ 57194.61 -- 57194.62 (included in ``H'' period in Fig. \ref{fig:lc}). }
\label{fig:sed}
\end{figure}
%%%%%%%%%%%%%%%%%%%%
To confirm our interpretations discussed above, 
we tried to describe the spectral energy distribution (SED) from optical to ultraviolet 
with a power-law and a irradiated disk spectra. 
Constructed SEDs at MJD 57194.616 (included in ``H'' portion in Fig.\ref{fig:lc}) are displayed in Fig.\ref{fig:sed}. 
Four {\it Swift}/UVOT fluxes (V, U, UVW1, and UVW2) are obtained quasi-simultaneously with the optical bands within 12 mins. 
We ignored the flux of the stable component since its contribution is insignificant 
relative to LVC and HVC ($\gtrsim$ 10 times lower than that of the minimum flux we observed in these time windows). 
The solid line and the dashed line denote the irradiated disk spectrum 
(the outermost temperature T$ _{\rm out} = 5\times10^3$ K with emitting size $R = 4 \times 10^{5}$ R$_s$) and 
the power-law spectrum with an index $\alpha = 1.0$ respectively. 
Although we did not use the UVOT data to derive the spectral model, 
all data points appear to fit nicely with the two component model including the UV excess. 
Using the spectral analysis, 
we additionally confirmed the results from flux-flux plot (\S\ref{sec:dec}) 
that the optical variation with large amplitude are mostly attributed to HVC. 
When we fit the optical SED with the two component model with fixed disk temperature and 
fixed power-law index for each time window (A--J), 
normalizations of LVC was quite stable compared to those of the power-law components. 
This behaviors are consistent with the variability in ``small-fast wiggles'' and ``big up-and-down swings'', respectively. 
On the other hand, the SED at dim phases seems to be affected by the stable component, 
which shows a redder spectrum than LVC. 
These consistencies strongly supports our interpretation. 

Our result that the optical emission consists of two variable components and one stable component, 
has possibilities to give an interpretation for the characteristic flux variations between optical and X-ray 
pointed out by Kimura et al. (2016); 
although the amplitudes of the variations in X-ray and optical are generally correlated 
and it indicates that both X-ray and optical variations are reflected the same phenomena, 
the X-ray flux variations were always much larger than the optical ones. 
The smaller amplitude in the optical-band relative to X-ray 
is naturally interpreted by the ``bias'' (stable component) lying under the variable component (HVC and/or LVC) correlated with X-ray variation. 
In MJD $\sim$ 57193.63 (in ``F'' segment), for example, a rapid fading was observed simultaneously in optical and X-ray. 
However, its amplitude was quite different; $\times$1/100 in X-ray while $\times$1/5 in R$_{\mathrm{C}}$ band. 
The stable component contributes at least 2\% and as much as 22\% of the peak flux before the fading in the R$_{\mathrm{C}}$ band. 
We therefore can attribute the difference in fading amplitude of optical and X-ray to the stable component 
sustaining the optical flux. 

\section{Summary}
In this work, we found that the optical emission of V404 Cygni in 2015 outburst can be decomposed to 
three components: the little-variable component (LVC), the highly-variable component (HVC), and the stable component, 
based on the analysis of the flux-flux plot which have a break and then is well represented by broken-liner expression. 

The spectral indices of LVC and HVC on the color-color diagram 
indicate that LVC shows multi-temperature black body spectra with the disk with $T(r) \propto r^{-1/2}$, 
while HVC shows power-law spectra. 
The inner and outermost temperatures ($T_{\rm in/out}$) and the outermost radius of the irradiated disk are 
$T_{\rm in}\gtrsim 2.0 \times 10^4$ K, $T_{\rm out}\lesssim 6.5 \times 10^3$ K, 
and (3--5) $\times10^4$ Schwarzschild radii respectively. 
The gradual brightening trend of the light curves over MJD 57182--57194 ($\sim 0.6$ mag) is mostly attributed to this disk component. 
On the other hand, the HVC has a power-law index of 0.6--1.0. 
We proposed a non-thermal jet or thermal cyclo-synchrotron emission as its origin of the power-law component (HVC). 
For the origin of remaining stable component, we suggested a large scale disk or the companion star wind,
ob the basis of its stability and H$\alpha$ line contribution suggested by its prominence in the R$_{\rm C}$-band. 

The decomposition of the optical flux variation into the two variable components based on the flux-flux plot analysis 
is supported by the analysis of the optical-UV spectral energy distribution, 
which can be described by the combination of an irradiated disk and a power-law component. 
We further find that 
observed optical spectra at different epochs can be reconstructed by the two spectral components with different time variability, 
``small-fast wiggles''  of the disk and ``big-slow swings'' of the power-law component. 

Additionally, we showed that the difference in variation amplitudes in optical and X-ray, which are generally correlated, 
can be interpreted with the flux contribution from stable component; 
it lies under the variable component and sustains the optical flux.

\begin{ack}
We thanks to the referee Dr. Josep Mart\'i for helpful comments and suggestions that improved the paper. 
Part of this work was financially supported by the Grant-inAid 
for JSPS Fellows for young researchers (Y.T.) and by 
Ministry of Education, Culture, Sports, Science and Technology of Japan (MEXT), 
Grant-in-Aid No.14GS0211, 19047003 and 24103002 (N.K.). 
We acknowledge support for MITSuME Telescope at Akeno 
by the Inter-University Research Program of the Institute for Cosmic Ray Research, University of Tokyo. 
This work was supported by the Optical and Near-infrared Astronomy Inter-University Cooperation Program 
and the JSPS-NSF PIRE Program. 
\end{ack}

\clearpage
%%%%%%%%%%%%%%%%%%%%
%%%%%%% Table %%%%%%%%%
%%%%%%%%%%%%%%%%%%%%

%%%%%%%%%%%%%%%%%%%%
%\begin{landscape}
\begin{table*}[]
\centering
%\rotate
\begin{minipage}{\textwidth}
\caption{The information of V404 Cyg, reference stars, and photometric standard stars.}
\begin{tabular}{lllcccc}\hline \hline
Objects & & & Magnitude & & & \\
Name & Attribution & RA, Dec (J2000) & $g^{\prime}$ & R$_{\rm C}$ & I$_{\rm C}$ \\ \hline
V404 Cyg & Target & (20:24:03.83, +33:52:02.2) & & & \\ \hline
SA111-1925 & Standard & (19:37:28.62, +00:25:03.1) & 12.545 & 12.167 & 11.914\\
SA111-1965 & & (19:37:41.55, +00:26:50.9) & 12.272 & 10.468 & 9.591\\
SA111-1969 & & (19:37:43.29, +00:25:48.6) & 11.370 & 9.205 & 7.983\\
SA111-2039 & & (19:38:04.58, +00:32:12.8) & 13.064 & 11.656 & 10.967\\
SA111-2088 & & (19:38:21.25, +00:31:00.4) & 13.992 & 12.305 & 11.487\\
SA111-2093 & & (19:38:23.44, +00:31:25.7) & 12.812 & 12.168 & 11.771\\
SA112-223 & & (20:42:14.58, +00:08:59.7) & 11.606 & 11.151 & 10.877\\
SA112-250 & & (20:42:26.38, +00:07:42.4) & 12.314 & 11.778 & 11.455\\
SA112-275 & & (20:42:35.43, +00:07:20.2) & 10.488 & 9.258 & 8.689\\
SA112-805 & & (20:42:46.74, +00:16:08.4) & 12.087 & 12.023 & 11.948\\
SA112-822 & & (20:42:54.90, +00:15:01.9) & 12.028 & 10.991 & 10.489\\ \hline
ref1 & Reference & (20:24:07.17, +33:50:51.9) & $13.045 \pm 0.007$ & $12.416 \pm 0.004$ & $12.011 \pm 0.004$\\
ref2 & & (20:23:53.39, +33:52:24.2) & $13.800 \pm 0.007$ & $12.762 \pm 0.005$ & $12.249 \pm 0.008$\\
ref3 & & (20:24:08.87, +33:54:38.3) & $13.403 \pm 0.008$ & $12.743 \pm 0.006$ & $12.341 \pm 0.008$\\
ref4 & & (20:23:56.48, +33:48:17.2) & $12.564 \pm 0.008$ & $12.224 \pm 0.004$ & $11.948 \pm 0.005$\\
ref5 & & (20:24:25.13, +33:51:56.1) & $13.247 \pm 0.007$ & $12.854 \pm 0.006$ & $12.550 \pm 0.007$\\ \hline
\end{tabular}
\label{tab:ref}
%\end{center}
\end{minipage}
\end{table*}
%\end{landscape}
%%%%%%%%%%%%%%%%%%%%
%%%%%%%%%%%%%%%%%%%%
\begin{table*}[]
\begin{center}
\begin{minipage}{\textwidth}
\caption{Log of optical observations.}
\begin{tabular}{lcccc}\hline\hline 
MJD & Duration & Exposure & Number & Location\\ 
	& [minute] & [sec] & [frame] & 	\\ \hline 
57192.55102 -- 57192.56430 & 19.1 & 60 & 15 & Ishigaki \\
57192.56605 -- 57192.57936 & 19.2 & 30 & 30 & Ishigaki \\
57192.58012 -- 57192.69439 & 164.5 & 20 & 333 & Ishigaki \\
57192.70261 -- 57192.83388 & 189.0 & 10 & 572 & Ishigaki \\
57193.54567 -- 57193.60504 & 85.5 & 10 & 278 & Ishigaki \\
57193.60549 -- 57193.83348 & 328.3 & 5 & 887 & Ishigaki \\
57194.54989 -- 57194.83326 & 408.1 & 20 & 577 & Ishigaki \\
57194.55391 -- 57194.74250 & 271.6 & 30 & 101 & Akeno \\ 
\hline \hline
\label{tab:log}
\end{tabular}
\end{minipage}
\end{center}
\end{table*}
%%%%%%%%%%%%%%%%%%%%
%%%%%%%%%%%%%%%%%%%%
%\begin{landscape}
\begin{table*}[]
\centering
\begin{minipage}{\textwidth}
\caption{Derived parameters by broken-linear model fit to flux-flux plots. }
\vskip 5pt
\scalebox{0.75}{
\begin{tabular}{lp{1pt}ccccp{1pt}cccc}
\hline
\multicolumn{1}{l}{MJD} && 
\multicolumn{4}{c}{Broken-linear model ( $g^{\prime}$ and R$_{\rm C}$ ) } && 
\multicolumn{4}{c}{Broken-linear model ( $g^{\prime}$ and I$_{\rm C}$ ) } \\
\multicolumn{1}{c}{	} && 
\multicolumn{1}{c}{$k_{\rm 1}$} & \multicolumn{1}{c}{$C$} & 
\multicolumn{1}{c}{$F_{\rm br}$} & \multicolumn{1}{c}{$k_{\rm 2}$} && 
\multicolumn{1}{c}{$k_{\rm 1}$} & \multicolumn{1}{c}{$C$} & 
\multicolumn{1}{c}{$F_{\rm br}$} & \multicolumn{1}{c}{$k_{\rm 2}$} \\
\hline
%%% 57192 %%%
57192.6456 -- 57192.7012 (A) && 
$0.756\ ^{+0.012}_{-0.010}$ & $0.111\ ^{+0.007}_{-0.008}$ & $1.292\ ^{+0.098}_{-0.090}$ & $0.639\ ^{+0.007}_{-0.006}$ && 
$0.660\ ^{+0.011}_{-0.016}$ & $0.088\ ^{+0.011}_{-0.008}$ & $1.286\ ^{+0.157}_{-0.083}$ & $0.527\ ^{+0.006}_{-0.010}$\\
57192.7012 -- 57192.7423 (B) && 
$0.824\pm0.016$ & $0.081\pm0.004$ & $0.788\ ^{+0.070}_{-0.056}$ & $0.646\ ^{+0.003}_{-0.004}$ && 
$0.699\ ^{+0.014}_{-0.016}$ & $0.069\pm0.004$ & $0.794\ ^{+0.066}_{-0.051}$ & $0.525\pm0.003$\\
57192.7423 -- 57192.7928 (C) && 
$0.831\ ^{+0.014}_{-0.015}$ & $0.081\pm0.004$ & $1.249\ ^{+0.132}_{-0.109}$ & $0.637\ ^{+0.011}_{-0.013}$ && 
$0.760\ ^{+0.017}_{-0.020}$ & $0.058\ ^{0.009}_{-0.006}$ & $1.130\ ^{+0.121}_{-0.086}$ & $0.512\ ^{+0.010}_{-0.013}$\\
57192.7928 -- 57192.8339 (D) && 
$0.803\ ^{+0.042}_{-0.038}$ & $0.107\ ^{+0.022}_{-0.028}$ & $1.023\ ^{+0.069}_{-0.195}$ & $0.665\ ^{+0.009}_{-0.010}$ && 
$0.702\ ^{+0.048}_{-0.074}$ & $0.091\ ^{+0.042}_{-0.036}$ & $0.828\ ^{+0.723}_{-0.028}$ & $0.560\ ^{+0.007}_{-0.029}$\\
%%% 57193 %%%
57193.5459 -- 57193.6083 (E) && 
$ - $ & $ - $ & $ < 2.420 $ & $0.665 \pm 0.007$ && 
$ - $ & $ - $ & $ < 2.420 $ & $0.553 \pm 0.007$\\
57193.6150 -- 57193.6390 (F) &&  
$0.898\ ^{+0.020}_{-0.048}$ & $0.055\ ^{+0.040}_{-0.024}$ & $1.407\ ^{+0.207}_{-0.010}$ & $0.676\ ^{+0.004}_{-0.006}$ && 
$0.833\ ^{+0.027}_{-0.054}$ & $-0.015\ ^{+0.043}_{-0.028}$ & $1.397\ ^{+0.173}_{-0.044}$ & $0.556\pm0.005$\\
57193.6390 -- 57193.8335 (G) && 
$0.769\ ^{+0.003}_{-0.004}$ & $0.138\ ^{+0.008}_{-0.007}$ & $ > 2.324 $ & $ - $ && 
$0.666\pm0.004$ & $0.091\ ^{+0.009}_{-0.008}$ & $ > 2.324 $ & $ - $ \\
%%% 57194 %%%
57194.5820 -- 57194.6617 (H) && 
$0.829\ ^{+0.025}_{-0.016}$ & $0.011\ ^{+0.039}_{-0.060}$ & $2.596\ ^{+0.022}_{-0.064}$ & $0.601\ ^{+0.007}_{-0.008}$ && 
$0.737\ ^{+0.032}_{-0.026}$ & $-0.057\ ^{+0.053}_{-0.075}$ & $2.591\ ^{+0.026}_{-0.057}$ & $0.447\pm0.008$\\
57194.6617 -- 57194.6722 (I) && 
$0.834\ ^{+0.054}_{-0.039}$ & $0.128\ ^{+0.035}_{-0.070}$ & $1.836\ ^{+0.310}_{-0.129}$ & $0.627\ ^{+0.005}_{-0.012}$ && 
$0.730\ ^{+0.069}_{0.053}$ & $0.090\ ^{+0.048}_{-0.089}$ & $1.801\ ^{+0.308}_{-0.146}$ & $0.488\ ^{+0.005}_{-0.009}$\\
57194.7404 -- 57194.8330 (J) && 
$0.910\ ^{+0.023}_{-0.021}$ & $0.043\ ^{+0.032}_{-0.036}$ & $1.966\ ^{+0.052}_{-0.113}$ & $0.629\ ^{+0.014}_{-0.009}$ && 
$0.802\pm0.023$ & $0.018\pm0.035$ & $1.990\ ^{+0.006}_{-0.181}$ & $0.490\ ^{+0.021}_{-0.017}$\\
\hline
\end{tabular}
}
\label{tab:par_sum}
\end{minipage}
\end{table*}
%\end{landscape}
%%%%%%%%%%%%%%%%%%%%

%%%%%%%%%%%%%%%%%%%%
%%%%%%% Figure %%%%%%%%%
%%%%%%%%%%%%%%%%%%%%

\end{document}